\documentclass[journal]{IEEEtran}
\usepackage{amsfonts}

\usepackage[ruled,vlined]{algorithm2e}
\usepackage{cite}
\usepackage{threeparttable}

\ifCLASSINFOpdf
  \usepackage[pdftex]{graphicx}
\else
\fi
\usepackage{amsmath}
\usepackage{url}
\usepackage{hyperref}
\hypersetup{
    colorlinks=true,
    linkcolor=blue,
    filecolor=magenta,      
    urlcolor=cyan,
}

\begin{document}
%
% paper title
% Titles are generally capitalized except for words such as a, an, and, as,
% at, but, by, for, in, nor, of, on, or, the, to and up, which are usually
% not capitalized unless they are the first or last word of the title.
% Linebreaks \\ can be used within to get better formatting as desired.
% Do not put math or special symbols in the title.
\iffalse
\title{Training Deep Models Robust to Dataset Bias with Bias-regularized Learning and Domain Augmentation}
\fi

\title{
Improve Model Generalization and Robustness to Dataset Bias with Bias-regularized Learning and Domain-guided Augmentation}

%
%
% author names and IEEE memberships
% note positions of commas and nonbreaking spaces ( ~ ) LaTeX will not break
% a structure at a ~ so this keeps an author's name from being broken across
% two lines.
% use \thanks{} to gain access to the first footnote area
% a separate \thanks must be used for each paragraph as LaTeX2e's \thanks
% was not built to handle multiple paragraphs
%

\author{Yundong~Zhang,
        Hang~Wu,
        Huiye~Liu,
        Li~Tong,
        and~May~D~Wang,~\IEEEmembership{Senior~Member,~IEEE}%
        %and~Jane~Doe,~\IEEEmembership{Life~Fellow,~IEEE}% <-this % stops a space

\iffalse
\thanks{H. Wu is with the Department
of Biomedical Engineering, Georgia Institute of Technology, Atlanta,
GA, 30332 USA e-mail: (see http://www.michaelshell.org/contact.html).}% <-this % stops a space
\thanks{XX are with Anonymous University.}
\fi
}
\maketitle

\begin{abstract}
Deep Learning has thrived on the emergence of biomedical big data. However, medical datasets acquired at different institutions have inherent bias caused by various confounding factors such as operation policies, machine protocols, treatment preference and etc. As the result, models trained on one dataset, regardless of volume, cannot be confidently utilized for the others. In this study, we investigated model robustness to dataset bias using three large-scale Chest X-ray datasets: first, we assessed the dataset bias using vanilla training baseline; second, we proposed a novel multi-source domain generalization model by (a) designing a new bias-regularized loss function; and (b) synthesizing new data for domain augmentation. We showed that our model significantly outperformed the baseline and other approaches on data from unseen domain in terms of accuracy and various bias measures, without retraining or finetuning. Our method is generally applicable to other biomedical data, providing new algorithms for training models robust to bias for big data analysis and applications. Demo training code is publicly available\footnote{\label{myfootnote}https://github.com/ydzhang12345/Domain-Generalization-by-Domain-guided-Multilayer-Cross-gradient-Training}.
\end{abstract}

% Note that keywords are not normally used for peerreview papers.

% For peer review papers, you can put extra information on the cover
% page as needed:
% \ifCLASSOPTIONpeerreview
% \begin{center} \bfseries EDICS Category: 3-BBND \end{center}
% \fi
%
% For peerreview papers, this IEEEtran command inserts a page break and
% creates the second title. It will be ignored for other modes.
\IEEEpeerreviewmaketitle

\section{Introduction}
\begin{figure*}[!t]
\begin{center}
\includegraphics[width=0.8\textwidth]{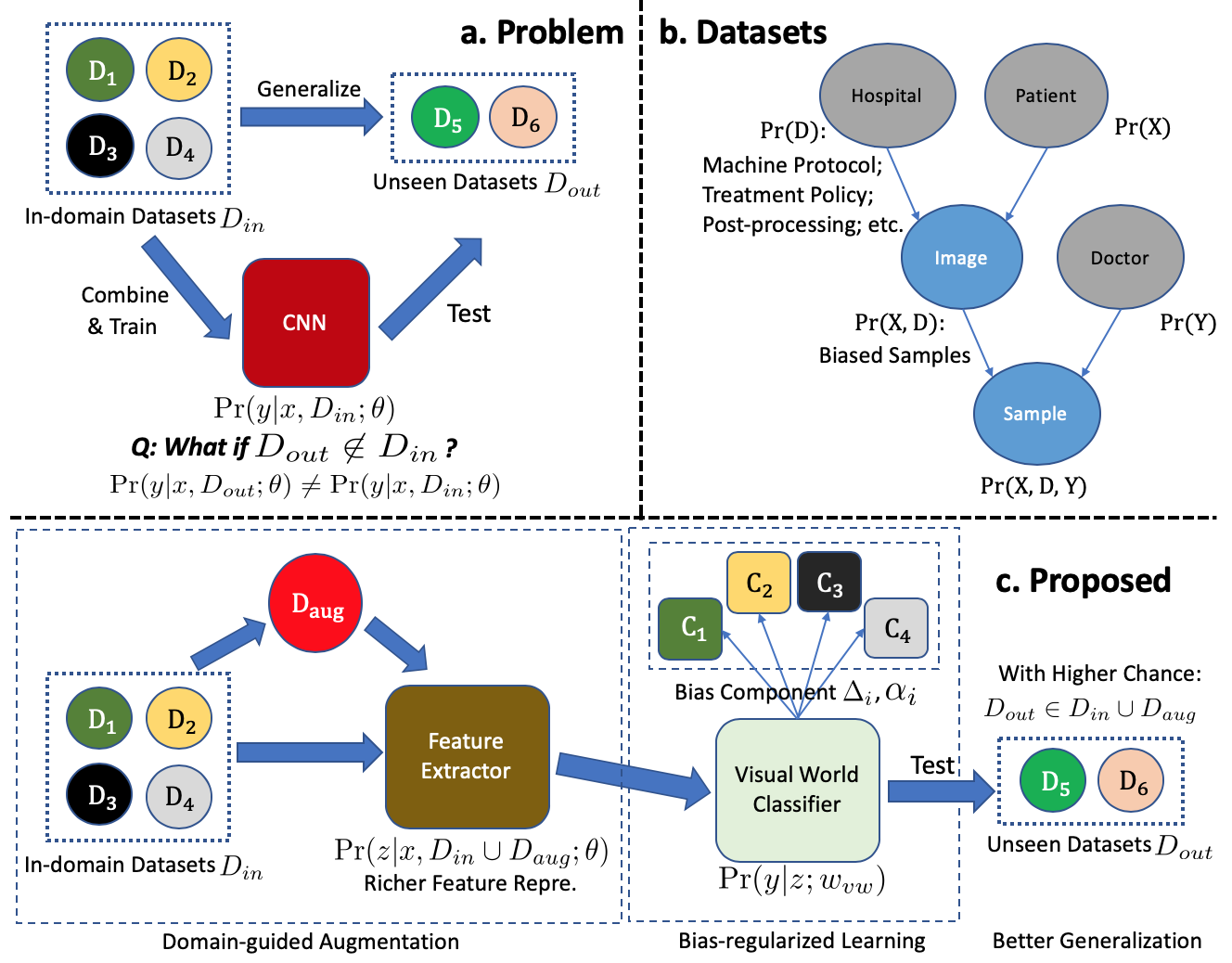}
%where an .eps filename suffix will be assumed under latex, 
% and a .pdf suffix will be assumed for pdflatex; or what has been declared
% via \DeclareGraphicsExtensions.    
\end{center}{}
\caption{(a) We are interested in learning under unknown dataset bias (domain shifts) problem, where test dataset can have different data distribution with the training sets; (b) a graph of common medical image dataset conditions, where the final observed samples are affected by hospital-specific factors, leading to dataset bias; (c) our proposed learning strategies, where we made two contributions: 1. Domain-guided augmentation for learning richer feature representation; 2. Bias-regularized Learning for promoting the domain-invariant features, where $\textup{C}_1$,...,$\textup{C}_4$ are the bias classifiers constructed from the visual world (debiased) classifier that is robust to dataset bias.}
\label{pull_fig}
\end{figure*}

\IEEEPARstart{D}ataset bias is inherent to big biomedical data, due to its heterogeneity, incidental endogeneity and dynamic nature \cite{fan2014challenges, ferryman2018fairness, lee2017medical}. Such bias mars the central assumption of machine learning models: both train and test data should be independent and identically distributed (IID). As the result, dataset bias leads to difference between the estimated and the true value of desired model parameters, which is the bias commonly used in statistics to describe the performance of an estimator \cite{walther2005concepts, west1999stereological}. Fig. \ref{pull_fig}a \& b describes the potential problems and sources of dataset bias in medical datasets, where $\textup{Pr}(D), \textup{Pr}(X)$ and $\textup{Pr}(Y)$ are the distribution of hospital-specific process, patients and doctors (labelers) respectively. Due to different hardware conditions, diagnosis policy and many more hospital-specific factors, $\textup{Pr}(D)$ is varying by hospitals. As a result, model trained on internal hospitals $\textup{Pr}(y|x,D_{in};\theta)$ cannot generalize to external sites $D_{out}$ if the internal datasets do not span the distribution of $\textup{Pr}(D_{out})$. This can be dangerous because clinicians might wrongly trust the system based on the internal observation.

\subsection{Case Study - Chest X-rays}
Recently, Zech et al. \cite{zech2018variable, zech_2018} collected in total 158,323 CXRs from three institutions (NIH, MSU and IU) and studied the dataset bias problem carefully. By manually reviewing the CXRs, they observed that: (1) More than 80\% of CXRs in MSU were tagged as portable inpatient scans, as those patients were too sick to move; in contrast, all CXRs of IU were outpatient and had no such tags on the images; as a result, model trained on MSU was leveraging the detection of portable tag to calibrate its prediction, leading to poor generalization in IU dataset; (2) NIH CXRs have chest tubes frequently for samples labelled with Pneumothoraxl; consequently, model from NIH heavily relied on the detection of obvious chest tubes instead of the subtler pneumothorax itself; this feature was not helpful for prediction because only after patients being treated would they have the chest tube. They further claimed that those hospital system-specific biases were complicated and hard to fully assess what factors were contributing the predictions exactly. But clearly all models were suffering from performance degradation on external set.

Following this line of research, we also collected more than 550,000 CXRs from three publicly available dataset: NIH, CheXpert and MIMIC \cite{wang2017chestx, irvin2019chexpert, johnson2019mimic}. Instead of manually studying the images, we use statistical tool called Classifier Test \cite{lopez2016revisiting} to unveil dataset biases. The basic idea is to train a dataset classifier to tell which dataset the input is from. If there is no domain shift or hospital-specific bias, the classifier should act as random guessing. Surprisingly, we found that despite data is of the same modality and body-part, a simple convolution neural network (CNN) achieved near-perfect classification accuracy (see Appendix \ref{conf_mat} for more details), meaning that severe dataset biases were induced during the creation of the image.
\iffalse
Additionally, by visualizing the classification heatmap (Fig. \ref{fig_name}a), we found that many regions con contribute to the dataset prediction, 
\fi

\iffalse
\begin{figure}[!t]
\begin{tabular}{cc}
\multicolumn{2}{c}{}\\
\includegraphics[width=0.2\textwidth]{name_the_dataset/heatmap.png} &
\includegraphics[width=0.23\textwidth]{name_the_dataset/name.pdf} 
\end{tabular}\\
%where an .eps filename suffix will be assumed under latex, 
% and a .pdf suffix will be assumed for pdflatex; or what has been declared
% via \DeclareGraphicsExtensions.    

\caption{Fig. \ref{fig_name}a. Activation heatmap for dataset classification, indicating the image corners contribute the most to the dataset classification; \ref{fig_name}b. confusion matrix of classification results. A perfect accuracy is obtained in \ref{fig_name}b, suggesting that dataset bias are present in the three datasets (MIMIC, CheXpert and NIH).}
\label{fig_name}
\end{figure}
\fi

\subsection{Problem Statement - Learning under Unknown Bias}
Developing useful machine learning models under intractable dataset bias has been well studied for the past few years \cite{torralba2011unbiased, tommasi2017deeper}. Without knowledge about the underlying mechanisms, most of the methods are trying to utilize datasets from multiple source to learn domain-invariant models \cite{khosla2012undoing, li2017deeper}. This is a common setting called multi-source domain generalization (MSDG). However, in practice we usually have rich samples of $\textup{Pr}(X)$ (data points) but very few $\textup{Pr}(D)$ (datasets). Thus, how to effectively utilize limited available datasets from different sources are the core of MSDG.

We here define our problem setting. Our goal is to train a model that perform well for in-domain datasets and generalize well to unseen domain. Formally, denote $D_i:=\{(x_1^{i}, y_1^{i}), (x_2^{i}, y_2^{i}),..., (x^{i}_{d_{i}}, y^{i}_{d_{i}})\} \in \mathcal{D}$, $i=1,...,N, N+1,...,M $ where $D_i$ is the $i^{th}$ dataset or sub-domain in a shared domain $\mathcal{D}$, $D_1,...D_N$ are the internal sets that are available during training and $D_{N+1},...,D_M$ are the external sets which are completely hidden unless on test time. Here we focus on the classification task and assume all the datasets share the common labels. Then we aim to 
\begin{equation} \label{goal}
    \max_{\Theta} \sum_{i=1}^N\sum_{j=1}^{d_i} s(\hat{y}_j^i, y_j^i|\Theta) + \sum_{i=N+1}^M\sum_{j=1}^{d_i} s(\hat{y}_j^i, y_j^i|\Theta)
\end{equation}{}
where $\hat{y}_j^i$ is the model prediction of sample $x_j^{i}$ parametrized by $\Theta$ and $s(\hat{y}_j^i, y_j^i)$ is our evaluation metric. The first double summation of \eqref{goal} is the internal set performance and the second part is that of external set. Notice that we have no access to the latter part of \eqref{goal} during training and hence our optimization can only focus on the former part.

\iffalse
In this work, we are specifically interested in bias of Biomedical Imaging. We observe that current advanced deep learning models are subject to dataset bias and suffer from generalization problem: an Alexnet model trained on large-scale CXRs dataset can exhibit about 10\% accuracy drop when testing on unseen data. Also, by looking at the feature embedding maps, we find that bias are learnt even without supervision (Fig. \ref{t-SNE-figure}). To close the generalization gap and mitigate the effect of dataset bias, we propose two strategies in the following sections.
\fi

\subsection{Contribution}
%To accommodate the biomedical data biases, we survey the classical machine learning literature and frame the problem as domain generalization where each dataset is a subset of a common domain.
To learn domain-invariant predictors more efficiently, we: 1. designed a new loss function that regularizes the bias learning by modelling each training domains explicitly; 2. proposed a new data augmentation methods to improve model generalization by generating domain-guided perturbed hidden activations (see Fig. \ref{pull_fig}). We verified our methods through extensive experiments and obtained superior performance compared to prior approaches.

For the following contents, we first introduced existing works on dataset bias and domain generalization in section \ref{related_work}; then in section \ref{methods}, we demonstrated the intuition and methodology of our proposed methods; as a case study, section \ref{exps} presented our experiments and provided quantitative and qualitative results on Chest X-ray datasets; we discussed our major findings and concluded the paper in \ref{discuss}.

\section{Related Work}\label{related_work}
Various metrics have been proposed to quantify the biases of dataset. The pioneering work in \cite{torralba2011unbiased} suggested use cross-dataset generalization: measure the relative performance drop between the original test set and the new dataset, as long as they come from the same domain. \cite{tommasi2017deeper} proposed to replace the relative measure by direct difference followed by a sigmoid function, in order to better preserve the information about internal test set. Cross-dataset generalization is an intuitive and interpretable measure, hence it is widely used in the machine learning community \cite{khosla2012undoing, mclaughlin2015data}. On the other hand, one can also perform Classifier Two-sample Test (C2ST) to verify whether two datasets are drawn from identical distribution \cite{lopez2016revisiting}. The idea is that if two datasets are in the same domain, a binary classifier trained on their joints should predict with chance-level on which dataset the sample is drawn. %The author use C2ST as a measurement to the quality of generative adversarial networks (GAN). In our work, we show that C2ST is an effective method for measure the dataset bias as well %On the other hand, \cite{li2018resound} and \cite{zhang2018examining} reports the issues of dataset representation biases, in which a complex dataset can be easily solved by some simple feature representation. Representation biases can be harmful when some spurious visual cues or text patterns become favored to represent the dataset by learning algorithms, while key semantic information is missed. \cite{li2018resound} use the log ratio between the performance of learning algorithm and the random guess to measure the representation biases of a particular feature on a given dataset. \cite{li2019repair} reformulates the cross entropy loss and define the representation biases as the reduction in uncertainty about the class label when a specific feature representation is used.
.

Besides the quantitative measures, several visualization techniques can be applied to qualitatively understand the source of biases. In \cite{khosla2012undoing}, the trained weights of a linear-SVM classifier were overlaid on original images to discover the pattern of their spatial distribution; \cite{zhou2016learning} generated the class activation heatmaps (CAM) to visualize the most contributed regions in input image for a trained CNN; \cite{selvaraju2017grad} proposed the guided backpropogation gradient activation heatmap (guided grad-CAM) to provide pixel-level attention of CNNs. In other fields such as natural language processing, the attention mechanism \cite{wang2016machine} is an effective way to visualize the focus of the model; One can also use Local Interpretable Model-Agnostic Explanations method (LIME) \cite{katuwal2016machine} to understand the model attentions. By comparing the model "attention" with human sense, we can verify whether the learning algorithm is learning the correct representation features and infer the source of data biases. If human attention heatmap is given, one can also use Spearman's rank correlation coefficients \cite{myers2013research} or earth mover's distance \cite{levina2001earth} to provide quantitative measures \cite{zhang2019interpretable, huk2018multimodal}. 

% above is measuring/visualizing dataset bias
Several framework has been developed to address dataset bias or domain generalization, where the goal is to train a model that generalizes to unseen datasets or domains. One of the earlier work \cite{khosla2012undoing} was based on max-margin learning (SVM), which modeled biases as per-dataset bias vectors in classifier space. During training, SVM maximized the objective of each dataset by constructing a classifier using addition of dataset-specific bias vector and bias-free vector; then for inference, the bias-free vector was used alone as bias-removal classifier. Built on top of it, \cite{tommasi2017deeper} conducted more extensive experiments by using DECAF features \cite{donahue2014decaf} as input to the model. The author found that the bias removal technique in \cite{khosla2012undoing} worked better when using classical BOWSift features \cite{DBLP:conf/iccv/Lowe99} while for DECAF features the opposite held. \cite{li2017deeper} further extended this shallow bias modelling structure to end-to-end training low-rank parametrized deep model and observed better performance.

Another series of work on domain generalization focus on feature level and aim to learn domain-invariant feature representation. In \cite{muandet2013domain}, a kernel-based method was used to project the data into common feature space where domain dissimilarity was minimized while the functional relationship of label was preserved. In \cite{Ghifary_2015_ICCV}, domain-robust feature was learnt by a multi-task data reconstruction autoencoders. Domain adversarial training technique could also be used for learning domain-independent feature by fooling a domain classifier \cite{ganin2016domain} or aligning distributions among different domains \cite{Li_2018_CVPR}.     %\cite{DBLP:journals/corr/abs-1811-12231} finds that almost all the popular models trained on ImageNet \cite{DBLP:conf/cvpr/2009} are strongly biased towards texture, which is contrast to human recognition that are usually based on object shapes. By introducing a new "stylized-ImageNet", \cite{DBLP:journals/corr/abs-1811-12231} improves the shape bias of ImageNet-trained CNNs and obtain better robustness as well as down-stream task performance (e.g. object detection). Following on this track, \cite{wang2019learning} propose a neural gray-level co-occurrence blocks and a projection-based mechanism that is able to force the model to "forget" some information. With these two novel methods, they are able to achieve comparable performance with respect to state-of-the-art domain adaptation techniques without using training samples from target domain. A similar work can be found in \cite{ganin2016domain}, where a reverse gradient method is applied to enforce two feature representations to be invariant to each other. \cite{li2019repair} also develop a new dataset resample paradigm that can improve the model generalization by training on a re-weighted dataset.

There are also efforts on addressing domain generalization through modifying the input data. \cite{carlucci2019domain} shuffled the original image patches and added an auxillary recompose task to the model to improve generalization. \cite{gan} used generative adversarial network (GAN) to generate domain-independent images. \cite{li2019repair} developed a new dataset resample paradigm (REPAIR) that improved the model generalization by training on a re-weighted dataset. The one most similar to ours is cross gradient training \cite{shankar2018generalizing}, which generated inter-domain data by domain-guided perturbations of the inputs.

A similar work that attempted to address dataset bias of Chest X-ray (CXR) data is \cite{yao2019strong}, where the authors collected ten CXR datasets internationally. However, they did not provide an effective method for handling dataset bias apart from directly trained and tested them in leave-one-out scheme. Also, their task (predict normal or abnormal) is simpler than ours.

\section{Proposed Methods}\label{methods}
\subsection{Bias-regularized Learning}
We start by revisiting the undoing-bias framework in \cite{khosla2012undoing}. Formally, let $z_j^i$ be the extracted feature of sample $x_j^i$, we aim to solve the following soft-constrained max-margin (SVM) optimization problem:
\iffalse
\begin{equation}\label{unbias}
\begin{split}
\min_{w_{vw}, \Delta_i, \xi,\rho}&\frac{1}{2}||w_{vw}^2|| + C_1\sum_{i=1}^{N}\sum_{j=1}^{d_i}\xi_{j}^{i} +\\ &\sum_{i=1}^{N}\left(\frac{\lambda}{2}||\Delta_i||^2  + C_2\sum_{j=1}^{d_i}\rho_j^{i}\right)
\end{split}{}
\end{equation}
%\begin{equation}
%    \min_{\textbf{w}_{vw}, \Delta_i, \xi,\rho, \theta_{vw}, \theta_i}\frac{1}{2}||\textbf{w}_{vw}^2|| + \frac{\lambda}{2}\sum_{i=1}^{n}||\Delta_i||^2 + C_1\sum_{i=1}^{n}\sum_{j=1}^{s_i}\xi_{j}^{i} + C_2\sum_{i=1}^{n}\sum_{j=1}^{s_I}\rho_j^{i}
%\end{equation}{}
\begin{align*}
\quad\quad\quad\mbox{s.t.}\quad &w_i = w_{vw} + \Delta_i, \\ 
&y_j^{i}w_{vw}\cdot z_j^{i} \ge 1 - \xi_j^{i}, \\ 
&y_j^{i}w_{i}\cdot z_j^{i} \ge 1 - \rho_j^{i}, \\
&\xi_j^{i} \ge 0, \rho_j^{i}\ge 0 , \\
& i=1,...,N \quad j=1,...,d_i
\end{align*}

where $d_i$ is the number of samples within dataset $D_i$; $\xi_j^i$ and $\rho_j^i$ are the slack variables to balance learning objectives; $\lambda$, $C_1$ and $C_2$ are hyperparameters; $\Delta_i$ is the bias vector of SVM; finally,  $w_{vw}$ is our desired de-biased classifier model. 
% needed in second column of first page if using \IEEEpubid
%\IEEEpubidadjcol
To optimize \eqref{unbias}, we can rewrite it into the soft-constrained form \cite{khosla2012undoing}
\fi
\begin{equation}\label{soft_un}
\begin{split}
    \min_{w_{vw},\Delta_i}&\frac{1}{2}||w_{vw}||^2+\frac{\lambda}{2}\sum_{i=1}^{N}||\Delta_i||^2\\+&\sum_{i=1}^{N}\sum_{j=1}^{d_i}C_1\max(1-y_j^i w_{vw} z_j^i,0)\\
    +&\sum_{i=1}^{N}\sum_{j=1}^{d_i}C_2\max\left(1-y_j^iw_iz_j^i,0\right)
\end{split}{}
\end{equation}{}
where $w_{vw}$ is our visual world (debiased) classifier, $w_i=w_{vw} + \Delta_i$ is the biased classifier for $i^{th}$ dataset and $C_1$, $C_2$ is the balancing hyper-parameters. Intuitively, here the learning of bias is controlled by regularizing the norm of classifiers. $w_{vw}$ is encouraged to only capture domain-invariant features while $w_i$ leverage dataset-specific features. 

The above SVM framework have several drawbacks: 1. the feature representation $z_j^i$ is not learnt end-to-end, where the original authors use BOWSift \cite{DBLP:conf/iccv/Lowe99}; 2. the hinge loss is not optimizer-friendly since it is not differentiable everywhere; 3. we cannot have a probability interpretation of the prediction, which is crucial in assisting medical diagnosis; 4. it models the bias weights with only additive relationl. Thus, we firstly propose to train the feature extractor end-to-end using deep model. That is, $z_j^i=F(x_j^i;\theta)$ where $F(\cdot;\theta)$ is a neural net feature extractor parametrized by $\theta$. During training, $\theta$ is updated by back-propagating the gradient through $z_j^i$. Secondly, 
we propose to train the network using cross-entropy loss to accommodate 1 and 2. Lastly, for 3 we introduce $\alpha_i$ for each dataset as an additional trainable parameters, such that $w_i=\alpha_i\odot w_{vw}+\Delta_i$, where $\odot$ represents the element-wise product. Here $\alpha_i$ models the multiplicative relation between the model bias and visual world. This enables the model to capture both the feature shifts and scaling of the bias datasets. With those changes, our proposed cross-entropy training objective is

\begin{equation}
\begin{split}
    L(\textbf{v})&=||w_{vw}||^2+\sum_{i=1}^{N} \lambda||\Delta_i||^2+\gamma||\alpha_i-1||^2 \\+&\sum_{i=1}^{N}\sum_{j=1}^{d_i}C_1 L_y(w_{vw}z_j^i, y_j^i) + C_2 L_y\left(w_iz_j^i, y_j^i\right)
\end{split}{}
\end{equation}{}
where $\textbf{v}=(w_{vw}, \theta, \Delta_i, \alpha_i)$ and $L_y$ is the negative log-likelihood between the last linear layer and ground-truth label.  

We highlight the importance of the regularization term $||w_{vw}||^2$ of visual world classifier in our proposed cross-entropy loss, because $w_{vw}$ can easily overfit on a solution that takes advantage of all the bias features, leading to poor generalization performance in external set. To see this, consider a learned feature embedding $f=[f_c, f_{b1}, f_{b2}]$, where $f_c$ is the common feature, $f_{b1}$ and $f_{b2}$ are the bias feature presented in $D_1$ and $D_2$. Ideally, we want $w_{vw}=[w_c, 0, 0]$ such that it can generalize to some unseen dataset or domain $D_3$. However, without proper regularization, $w_{vw}=[w_c, w_{1}, w_{2}]$ can still be a valid solution if for $D_1$ we have $f_1=[f_c, f_{b1}, 0]$ and for $D_2$ we have $f_2=[f_c, 0, f_{b2}]$. By penalizing the norm of $w_{vw}$ more than $\Delta_i$ and $(\alpha_i-1)$, we push the bias learning to those bias vectors instead of visual world classifier.  

\subsection{Domain-guided Augmentation}
The above framework model the bias in higher level classifier space. However, since we have limited training domains, our feature extractor can still overfit on the in-domain data. Synthesizing new domain data and augment them to training set could be a potential solution, as the model could have higher chance to span target domain distribution. One simple strategy of augmenting unseen domain data is to use Mix-up \cite{zhang2017mixup, verma2018manifold} technique, where synthetic data are generated by linearly mixing samples from different datasets. However, as we will show in the following section, when the dataset-bias are severe, this method suffer from the convergence problem. Also, the diagnosis of medical imaging are usually relying on fine-grained features of the image. In this case, naively mixing samples will destroy the crucial details in the data.

Another idea is to augment the training data by Cross-gradient Training method \cite{shankar2018generalizing}. Formally, consider a dataset classifier $g_d(\cdot;\phi)$ and data point $(x^i_j, y^i_j)$, we can generate a new sample $\bar{x}^i_j=x^i_j + \epsilon\nabla_{x^i_j}L_d(g_d(z^i_j;\phi), i)$, where $\epsilon$ is the step size and $L_d$ is the dataset classification loss. Then we can train the model with this synthetic data $(\bar{x}^i_j, y_i^j)$. To ensure the gradient change have minimum effect on the label $y_i^j$, we also augment the training of $g_d(\cdot;\phi)$ with $(\tilde{x}^i_j, i)$, where $\tilde{x}^i_j=x^i_j + \epsilon\nabla_{x^i_j}L_y(w_{vw}z^i_j, y^i_j)$. This makes the dataset classifier unsensitive to the labels and hence $\nabla_{x^i_j}L_d(g_d(z^i_j;\phi), i)$ won't change $y_j^i$. 

\begin{figure}[t]
\begin{center}
\includegraphics[width=0.35\textwidth]{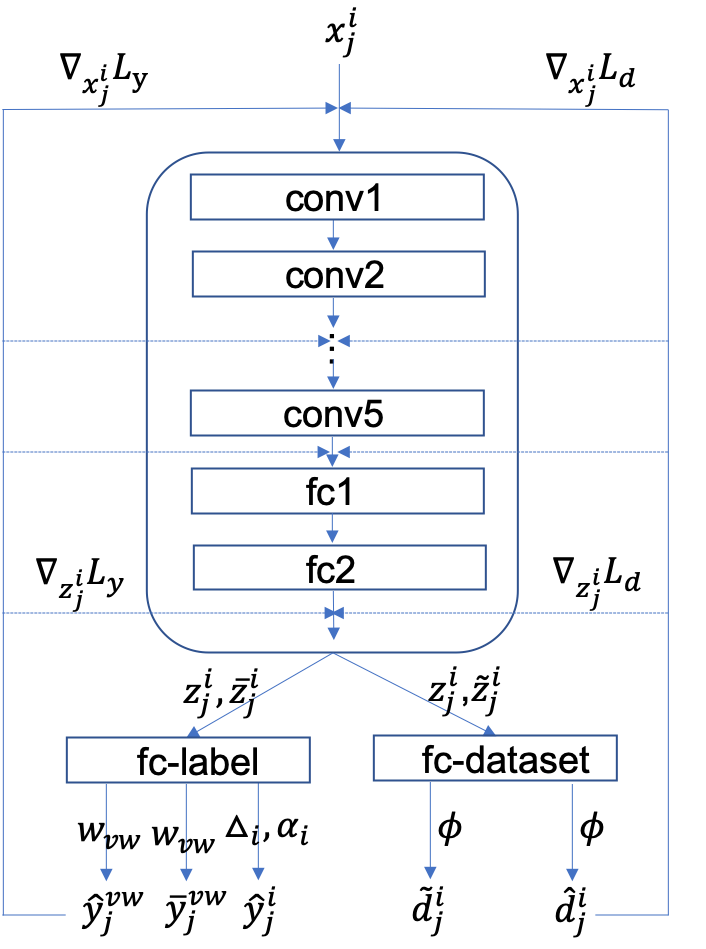}
\end{center}{}
\caption{Model Structure of our proposed MCT with Classifier-bias Modelling. Here $\hat{y}_j^i$ is the bias classifier prediction of original sample $j$ of dataset $i$; $\hat{y}_j^{vw}$, $\bar{y}_j^{vw}$ are the visual world classifier predictions of original sample and domain-guided augmented sample; $\hat{d}_j^i$, $\tilde{d}_j^i$ are the domain classifier prediction of original sample and augmented sample. During training, a layer activation (e.g. $z^i_j$) is randomly chosen from pre-determined set $\mathbb{S}$ and its domain-guided pertubation versions ($\hat{z}_j^i$ and $\tilde{z}_j^i$) are generated to augment the training.}
\vspace{-10pt}
\label{fig:model_fig}
\end{figure}

Our proposed method is built upon this idea. However, differ to \cite{shankar2018generalizing}, we do not train a separate network for $g_d$. Instead, $g_d$ is just a linear layer which directly takes the feature embedding $z_j^i=F(x_j^i;\theta)$ as input and output the dataset classification logits. During training, the gradient $\nabla_{x^i_j}L_d(g_d(z^i_j;\phi), i)$ will not propagate through $F(\cdot;\theta)$ and $L_d$ is only used to update $\phi$. The reasons for this design lie in two folds: training a separate feature extractor for $g_d$ or propagating the gradient through $\theta$ will lead to gradient vanishing $||\nabla_{x^i_j}L_d(g_d(z^i_j;\phi),i)||\approx0$, because the dataset classifier is significantly easier to train; more importantly, observing that now $L_d$ is differentiable with respect to $\theta$, we can augment the intermediate features in addition to the input $x_j^i$, leading to a new multi-layer augmentation training paradigm. Specifically, given a pre-determined set of layer output, e.g. $\mathbb{S}=(x^i_j, q^i_j, z^i_j)$, where $q^i_j$ and $z^i_j$ are the output of the last convolution and fully-connected layer, respectively. For each training step we can sample one of the layer output, say $q^i_j$, compute $\nabla_{q^i_j}L_d(g_d(z^i_j;\phi),i)$, generate a new augmentation feature point $\bar{q}_j^i=q^i_j + \epsilon\nabla_{q^i_j}L_d(g_d(z^i_j;\phi), i)$ and feed $\bar{q}_j^i$ to the following layers. We shall also follow Cross-gradient Training to generate $\tilde{q}_j^i=q^i_j + \epsilon\nabla_{q^i_j}L_y(w_{vw}z^i_j, y^i_j)$ and augment the training of $L_d$. Because the augment data point does not belong to any specific datasets, $\hat{q}^i_j$ is only fed to the visual world classifier for training $\theta$ and $w_{vw}$. In this way, we improve the robustness of our feature extractor to unseen data. 
\iffalse
In sum, our final training objective is
\begin{equation}
\begin{split}
    L(\textbf{v})&=||w_{vw}||^2+\sum_{i=1}^{n} \lambda||\Delta_i||^2+\gamma||\alpha_i-1||^2 \\&+\sum_{i=1}^{n}\sum_{j=1}^{m_i}C_1 L_y(w_{vw}F(z_j^i, y_j^i) + C_2 L_y\left(w_iz_j^i, y_j^i\right)\\
    &+C_3 L_y(w_{vw}p_j^i,y_j^i)
\end{split}{}
\end{equation}{}
\fi
We name this novel domain-guided augmentation method as \textit{Multi-layer Cross-gradient Training (MCT)}. Together with the Bias-regularized Learning, we summarize the overall model pipeline in Fig. \ref{fig:model_fig} and the training pseudo-code in Algorithm \ref{Algorithm}. 

%\iffalse
\begin{algorithm*}[h]
\KwIn{Multi-source labeled data $D_i={(x_1^i,y_1^i), (x^i_2,y^i_2),...,(x_d^i,y_d^i)}, i=1,...,N$; step size $\epsilon$, learning rate $\eta$, number of training steps $s$, trade-off parameters $C_1$, $C_2$, $C_3$, $\lambda$, $\gamma$; selected augmentation layer set $\mathbb{S}$.}
\KwOut{Label and dataset classifier parameters $\theta$,$\phi$,$w_{vw}$,$w_i$,$\Delta_i$, $\alpha_i$, $i=1,...,N$.}
Randomly initialize $\theta$,$\phi$,$w_{vw}$ and $\Delta_i$, initialize $\alpha_i=1$, $i=1,...,N$.

\SetKwFunction{FMain}{MCT}
\SetKwProg{Fn}{Function}{:}{}
\Fn{\FMain{label loss $L_y$, domain loss $L_d$, layer set $\mathbb{S}$}}{
Randomly sample a layer from $\mathbb{S}$ and denote the sampled layer embedding as Q;\\
Generate augmentation sample $\bar{Q}=Q + \epsilon\nabla_{Q}L_d$ and $\tilde{Q}=Q + \epsilon\nabla_{Q}L_y$;\\
Feed $\bar{Q}$ and $\tilde{Q}$ to the following layers, obtaining augmented final embedding $\bar{Z}$ and $\tilde{Z}$;\\
\textbf{return} $\bar{Z}$,  $\tilde{Z}$
}
\textbf{End Function}

\For{s \textup{training steps}}{
Randomly sample a mini-batch $(X, Y, D)$, where $D$ contains the dataset labels of each sample;\\
Compute the feature embedding $Z=F(X;\theta)$, visual world label classification loss $L_{y}(w_{vw}Z, Y)$ and the dataset classification loss $L_{d}(g_d(Z;\phi),D)$;\\
Call function \texttt{MCT}$\left(L_{y}(w_{vw}Z, Y), L_{d}(g_d(Z;\phi),D), \mathbb{S}\right)$ to generate augmented embedding $\bar{Z}$ and $\tilde{Z}$;\\
Compute the augmentation training loss $L_y(w_{vw}\bar{Z}, Y)$ and $L_d(g_d(\tilde{Z};\phi),D)$;\\
Split $(X,Y,D)$ into per-dataset batch $(X^i, Y^i, i)$; compute bias classifier loss $L_y(w_iZ,Y)$;\\
 Update $\theta=\theta - \eta\nabla_{\theta}\left(C_1L_y(w_{vw}Z, Y)+C_2L_y(w_iZ, Y)+C_3L_y(w_{vw}\bar{Z}, Y)\right)$;\\
 Update $w_{vw}=w_{vw} - \eta\nabla_{w_{vw}}\left(C_1L_y(w_{vw}Z, Y)+C_2L_y(w_iZ, Y)+C_3L_y(w_{vw}\bar{Z}, Y) + ||w_{vw}||^2\right)$;\\
 Update $\Delta_i=\Delta_i - \eta\nabla_{\Delta_i}\left(C_2L_y(w_iZ, Y)+\lambda||\Delta_i||^2\right)$; similarly to $\alpha_i$;\\
 Update $\phi = \phi - \eta\nabla_{\phi}\left(L_d(g_d(Z;\phi), D)+L_d(g_d(\tilde{Z};\phi), D)\right)$;}
\textbf{End for}
\caption{{\bf MCT with Bias-regularized Learning} \label{Algorithm}}
\end{algorithm*}

\section{Experiments}\label{exps}
\subsection{Datasets}
We use three large-scale Chest X-ray datasets as a case study, which are all open-sourced. They are NIH ChestX-ray14 from  NIH Clinical Center \cite{wang2017chestx}, Stanford CheXpert from Stanford Hospital \cite{irvin2019chexpert} and Mimic-CXR from Beth Israel Deaconess Medical Center \cite{johnson2019mimic, goldberger2000physiobank}. Since the above datasets have different label categories, we select 5 common diseases (Atelectasis, Cardiomegaly, Consolidation, Edema and Effusion) that they share with each other. We also discard all the lateral scans in CheXpert and Mimic-CXR as NIH only have frontal view images. Table \ref{table_example2} summarizes the basic information of each processed dataset. We use a roghly 7:1:2 split for train, val and test set of each dataset, except for NIH which has an official split that has the same ratio. We also ensure X-ray scans of the same patient belong to the same split set, preventing information leakage. 

We also include three popular datasets of domain generalization to specifically verify our proposed Multi-layer Cross-gradient Training algorithm. Dataset details and experiments can be found in Appendix \ref{toy exp}. 

\begin{table*}[!t]
% increase table row spacing, adjust to taste
\renewcommand{\arraystretch}{1.3}
% if using array.sty, it might be a good idea to tweak the value of
% \extrarowheight as needed to properly center the text within the cells
\caption{Summary of Three CXR Datasets}
\label{table_example2}
\centering
%% Some packages, such as MDW tools, offer better commands for making tables
%% than the plain LaTeX2e tabular which is used here.
\begin{tabular}{|c|c|c|c|c|c|c|c|}
\hline
 Datasets & \# Patients & \# Scans & \# Atelectasis & \# Cardiomegaly & \# Consolidation & \# Edema & \# Effusion \\
\hline
NIH & 30806 & 112120 & 11559 & 2776 & 4667 & 2303 & 13317  \\
\hline
CheXpert & 64534 & 191027 & 59583 & 23385 & 12983 & 61493 & 76899  \\
\hline
MIMIC & 62592& 248236 & 60681 & 48894 & 11733 & 43559 & 58731\\
\hline
\end{tabular}
\end{table*}

\subsection{Bias Measurements Metrics}
In this section we introduce two quantitative metrics for measuring biases of trained models.
\subsubsection{Generalization-Based Metrics}
The generalization-based metric works by evaluating how the model performs when trained on internal sets and test on external sets. Following \cite{torralba2011unbiased, tommasi2017deeper}, let
\begin{equation}
    self:=\frac{1}{N}\sum_{i=1}^N\frac{1}{d_i}\sum_{j=1}^{d_i}s(\hat{y}_j^i, y_j^i)
\end{equation}
be the internal test set performance and 
\begin{equation}
    others:=\frac{1}{M-N}\sum_{i=N+1}^M\frac{1}{d_i}\sum_{j=1}^{d_i}s(\hat{y}_j^{i}, y_j^{i})
\end{equation}{}
be the average external test set performance. The cross-dataset performance drop $PD$ for a particular model can be defined as:
\begin{equation}
    PD = \frac{self - others}{self}
\end{equation}

Intuitively, $PD$ measures the change of cross dataset performance, normalized by the internal set score. $PD > 0$ indicates that biases are present, which becomes more severe when it gets closer to 1. If $PD < 0$, it means internal performance is sub-optimal and no informative conclusion can be drawn by cross-dataset evaluation.

\subsubsection{Classifier Two-sample Test}
The Classifier Two-sample Test (C2ST) aims to determine whether two datasets are drawn from the same distribution by training a binary classifier to differ from each other. Formally, given two datasets $D_{1}$ and $D_{2}$, we can construct a new dataset \cite{lopez2016revisiting}
\begin{equation}
    D_{te} = \{(x^{1}_i, 0)\}_{i=1}^{d_{1}} \cup \{(x^{2}_i, 1)\}_{i=1}^{d_2}:=\{(z_i, l_i)\}_{i=1}^{d_{te}}
\end{equation}
and a binary classifier $g:\mathcal{X}\xrightarrow{}[0,1]$ to be the conditional probability estimation of $p(l_i=1|z_i)$. Then we can obtain the classification accuracy according to
\begin{equation}\label{acc}
Acc = \sum_{i=1}^{d_{te}}\mathbb{I} \left[ \mathbb{I}\left( g(z_i) > 0.5\right)=l_i\right].
\end{equation}{}
Intuitively, if the two datasets are from the same distribution, the test set $Acc$ should be close to 0.5, i.e. no better than random guessing. Otherwise, there must be some distinct features in one of the dataset which are exploited by the classifier.

\subsection{Results on Large-scale CXRs}
In this section, we evaluate our proposed methods on Large-scale CXR datasets. For training, we resize all the images to 256$\times$256, followed by a random crop of $224\times224$. Unlike \cite{irvin2019chexpert}, we do not use random horizontal flip since some diseases (e.g. Cardiomegaly) rely on spatial information. AlexNet \cite{krizhevsky2012imagenet} pretrained on Imagenet \cite{deng2009imagenet} is used as feature extraction backbone for all the models which are compared in the following sections unless specified. As each patient can have multiple diseases at the same time, our task is essentially a multi-label classification. We use binary cross-entropy loss for each diseases. Hyperparameters are determined by validation set and the selected $\mathbb{S}$ of MCT here is the input and the last dense layer of feature extraction network.  We implement all our experiments with Pytorch \cite{paszke2017automatic}.

\iffalse
\subsubsection{Name the Dataset}
We first perform Name the Dataset study as in \cite{torralba2011unbiased} and \cite{tommasi2017deeper}. In this task, we build a simple 3-layer CNN to classify input image into one of the three collected datasets. Fig. \ref{fig_name} shows the classification result on a random subset. Surprisingly, we find that despite these three datasets belong to the same modality and scanning on the same parts of human body, nearly perfect classification accuracy is obtained, meaning that severe dataset biases are induced during the creation of the final image.

\begin{figure}[!t]
\begin{center}
\includegraphics[width=0.4\textwidth]{name_the_dataset/name.pdf}
%where an .eps filename suffix will be assumed under latex, 
% and a .pdf suffix will be assumed for pdflatex; or what has been declared
% via \DeclareGraphicsExtensions.    
\end{center}{}
\caption{Confusion Matrix of Name that Dataset Experiments.}
\label{fig_name}
\end{figure}
\fi

\subsubsection{Classification on Seen Datasets}
Remind that we not only want our model generalize well on unseen sets, but also on the available internal sets. Thus, we first evaluate how our proposed methods perform on seen datasets. We use leave-one-out scheme for splitting the domains and run experiments on all possible dataset combinations. Table \ref{table:internal} shows the internal performance of various models.

\begin{table*}[!t]
% increase table row spacing, adjust to taste
\renewcommand{\arraystretch}{1.3}
% if using array.sty, it might be a good idea to tweak the value of
% \extrarowheight as needed to properly center the text within the cells
\caption{AUCS Score of Different Models on Internal Set of CXR datasets}
\label{table:internal}
\centering
%% Some packages, such as MDW tools, offer better commands for making tables
%% than the plain LaTeX2e tabular which is used here.
\begin{threeparttable}
\begin{tabular}{|c|c|c|c|}
\hline
 Models & CheXpert+MIMIC$\rightarrow$NIH & CheXpert+NIH$\rightarrow$MIMIC & MIMIC+NIH$\rightarrow$CheXpert \\
\hline
Alexnet & 0.817$\pm$0.002 & 0.811$\pm$0.001 & 0.831$\pm$0.001  \\
%\hline
DANN & 0.758$\pm$0.003 & 0.789$\pm$0.002 & 0.806$\pm$0.003  \\
%\hline
RAPAIR & 0.812$\pm$0.002 & 0.808$\pm$0.000 & 0.829$\pm$0.001\\
Mixup & 0.805$\pm$0.002  & 0.809$\pm$0.001 & 0.825$\pm$0.002 \\
CrossGrad & 0.816$\pm$0.001 & 0.811$\pm$0.000 & 0.831$\pm$0.000 \\
\hline
%HexBias & 0.761 & 0.828 & 0.766 & 0.857 & 0.890 & 0.820 \\
%\hline
E2E-SVM(bias) & 0.812$\pm$0.001(0.815$\pm$0.000) & 0.801$\pm$0.002(0.809$\pm$0.001) & 0.822$\pm$0.002(0.828$\pm$0.001)  \\
%\hline
E2E-CE(bias) & 0.816$\pm0.000$(0.819$\pm$0.001) & 0.804$\pm$0.000(0.814$\pm$0.000) & 0.823$\pm$0.000(0.833$\pm$0.000)  \\
%\hline
E2E-CE+Cg(bias) & 0.816$\pm0.001$(0.818$\pm$0.001) & 0.804$\pm$0.000(0.814$\pm$0.001) & 0.823$\pm$0.001(0.832$\pm$0.000) \\ 
%\hline
%Ours-CG-vw(bias) & 0.757(0.759) & 0.812(0.827) & 0.766(0.768) & 0.855(0.857) & 0.888(0.889) & 0.816(0.820) \\
\textbf{E2E-CE+MCT(bias)} & \textbf{0.817$\pm$0.001(0.820$\pm$0.000)} & \textbf{0.805$\pm$0.000(0.814$\pm$0.000)} & \textbf{0.824$\pm$0.000(0.833$\pm$0.000)}  \\
\hline
\end{tabular}

\begin{tablenotes}
\footnotesize
\item For the last four models where we have bias classifier $w_i$ and visual world classifier $w_{vw}$, we show AUC score of both and the one in bracket is the bias result  %此处加入注释*信息
        %\item[**] my website is ... %此处加入注释**信息
\end{tablenotes}
\end{threeparttable}
\end{table*}

We choose vanilla Alexnet as our baseline and also compare our methods to several advanced models on domain adaptation and generalization. Specifically, in the baseline we combine and train all the datasets together (train-them-all-together) without other special strategy. In domain adversarial training (DANN) \cite{ganin2016domain} we want to learn a domain-invariant feature representation by fooling a dataset discriminator. In REAPIR \cite{li2019repair}, we fix a trained feature extractor and assign a trainable weight for each sample to minimize the dataset representation bias; then we resample the dataset according to the weights and retrain the network. In Mixup \cite{zhang2017mixup}, inputs and labels are modified to be weighted sum of data from different domains. In CrossGrad \cite{shankar2018generalizing}, adversarial inputs are synthesized guided by domain perturbations. We found that DANN fails to converge because the gradient of domain discriminator dominates the feature learning; REPAIR does not help for the internal performance; Mixup is worse than vanilla baseline; Crossgrad suffers from gradient vanishing problem. On the contrary, our proposed undoing bias framework with cross-entropy loss (E2E-CE) surpass all the alternatives, suggesting that there is performance gain by modelling dataset bias carefully in multi-source data training. The performance is further increased by using our proposed MCT augmentation. Notice that the bias weight vector in our proposed model performs better than the visual world ones in the internal set, indicating that our model effectively encode dataset-specific information in the bias model.
\subsubsection{Classification on unseen dataset} 
Table \ref{table:external} demonstrates the AUC score of each model tested on external set. Unlike what is found in \cite{tommasi2017deeper} where the undoing bias framework \cite{khosla2012undoing} perform worse with DECARF feature, we show that by training the model end-to-end we can in fact get better performance on external generalization. Moreover, we observe similar results as in internal set performance. Our proposed method surpass all the comparing methods in every domain split, closing the performance gap between internal and external domain. We also find that popular domain adaptation methods such as DANN \cite{ganin2016domain} and data augmentation methods such as Mixup \cite{zhang2017mixup} do not work well for CXR data. 

\iffalse
\begin{table*}[!t]
% increase table row spacing, adjust to taste
\renewcommand{\arraystretch}{1.3}
% if using array.sty, it might be a good idea to tweak the value of
% \extrarowheight as needed to properly center the text within the cells
\caption{AUCS Score of Different Models on External Set of CXR datasets}
\label{table:external}
\centering
%% Some packages, such as MDW tools, offer better commands for making tables
%% than the plain LaTeX2e tabular which is used here.
\begin{tabular}{|c|c|c|c|c|c|c|c|}
\hline
 Models & Atelectasis & Cardiomegaly & Consolidation & Edema & Effusion & Average \\
\hline
Alexnet & 0.697 & 0.758 & 0.669 & 0.786 & 0.789 & 0.740 \\
%\hline
DANN \cite{ganin2016domain} & 0.643 & 0.694 & 0.678 & 0.758 & 0.754 & 0.705 \\
%\hline
REPAIR \cite{li2019repair} & 0.702 & 0.768 & 0.670 & 0.779 & 0.789 & 0.741\\
%\hline
Mixup \cite{zhang2017mixup} & 0.676 & 0.738 & 0.674 & 0.785 & 0.768 & 0.728\\
%\hline
CrossGrad \cite{shankar2018generalizing} & 0.701 & 0.760 & 0.670 & 0.783 & 0.790 & 0.741\\
\hline
%HexBias & 0.713 & 0.798 & 0.669 & 0.784 & 0.800 & 0.753 \\
%\hline
E2E-SVM-vw & 0.703 & 0.782 & 0.675 & 0.788 & 0.789 & 0.747 \\
%\hline
E2E-CE-vw & 0.707 & 0.793 & 0.677 & 0.791 & 0.793 & 0.752 \\
%\hline
\textbf{Ours} & \textbf{0.712} & \textbf{0.796} & \textbf{0.681} & \textbf{0.795} & \textbf{0.798} & \textbf{0.756} \\
\hline
\end{tabular}
\end{table*}
\fi

\begin{table*}[!t]
% increase table row spacing, adjust to taste
\renewcommand{\arraystretch}{1.3}
% if using array.sty, it might be a good idea to tweak the value of
% \extrarowheight as needed to properly center the text within the cells
\caption{AUCS score of different models for common chest diseases on external set}
\label{table:external}
\centering
%% Some packages, such as MDW tools, offer better commands for making tables
%% than the plain LaTeX2e tabular which is used here.
\begin{tabular}{|c|c|c|c|}
\hline
 Models & CheXpert+MIMIC$\rightarrow$NIH & CheXpert+NIH$\rightarrow$MIMIC & MIMIC+NIH$\rightarrow$CheXpert \\
\hline
Alexnet & 0.740 $\pm$ 0.002 & 0.800$\pm$0.001 & 0.756$\pm$0.000  \\
%\hline
DANN & 0.705$\pm$0.006 & 0.788$\pm$0.003 & 0.723$\pm$0.005  \\
%\hline
RAPAIR & 0.741$\pm$0.001 & 0.799$\pm$0.002 & 0.757$\pm$0.001\\
Mixup & 0.735$\pm$0.002  & 0.797$\pm$0.000 & 0.755$\pm$0.001 \\
CrossGrad & 0.742$\pm$0.000 & 0.801$\pm$0.000 & 0.755$\pm$0.001 \\
\hline
%HexBias & 0.761 & 0.828 & 0.766 & 0.857 & 0.890 & 0.820 \\
%\hline
E2E-SVM & 0.748$\pm$0.002 & 0.801$\pm$0.002 & 0.758$\pm$0.001  \\
%\hline
E2E-CE & 0.752$\pm$0.001 & 0.805$\pm$0.000 & 0.761$\pm$0.001  \\
%\hline
E2E-CE+Cg & 0.752$\pm$0.001 & 0.804$\pm$0.001 & 0.760$\pm$0.002  \\
%\hline
%Ours-CG-vw(bias) & 0.757(0.759) & 0.812(0.827) & 0.766(0.768) & 0.855(0.857) & 0.888(0.889) & 0.816(0.820) \\
\textbf{E2E-CE+MCT} & \textbf{0.755$\pm$0.000} & \textbf{0.807$\pm$0.001} & \textbf{0.763$\pm$0.001}  \\
\hline
\end{tabular}
\end{table*}

\subsubsection{Measuring The Bias}
In this section, We further present the bias measurements of each model. We extract the hidden representation of trained models at the last layer of feature extractor and train a dataset classifier on top of that to perform the classifier Test. Table \ref{table:bias_measure} summarizes the results. Several observations can be drawn: 1. although CXR data seems to be very similar to each other regardless of its origin, training a deep model naively can significantly induce dataset bias, leading to large generalization gap when testing on other source of data;  2. our proposed methods obtain the best average performance with much smaller performance drop between internal set and external set; 3. our proposed methods have smaller dataset bias.

\begin{table}[!t]
% increase table row spacing, adjust to taste
\renewcommand{\arraystretch}{1.3}
% if using array.sty, it might be a good idea to tweak the value of
% \extrarowheight as needed to properly center the text within the cells
\caption{Model Bias Comparison}
\label{table:bias_measure}
\centering
%% Some packages, such as MDW tools, offer better commands for making tables
%% than the plain LaTeX2e tabular which is used here.
\begin{tabular}{|c|c|c|c|}
\hline
 Models & Perf. Drop (PD) & Classifier Test & Rank Cor. \\
\hline
Alexnet  & 9.42\% & 99.37\% & 0.05   \\
\hline
DANN  & 6.99\% & 77.58\% & 0.02 \\
\hline
REPAIR & 8.74\% & 98.21\% & 0.04 \\
\hline
%HexBias & ? & 8.17\% & 99.?\% & \textbf{0.17} & ? \\
%\hline
%E2E-SVM-vw & 7.88\% & 91.33\% & 0.10  \\
%\hline
%E2E-CE-vw & 7.84\% & 94.56\% & 0.09  \\
%\hline
CrossGrad & 9.07\% & 97.49\% & 0.07\\
\hline
E2E-CE+MCT & 7.59\% & 92.86\% & 0.12  \\
\hline
\end{tabular}
\end{table}

\begin{figure}[h!]
\begin{center}
\begin{tabular}{cc}
\multicolumn{2}{c}{}\\%t-SNE Visualization}\\
\includegraphics[width=0.2\textwidth]{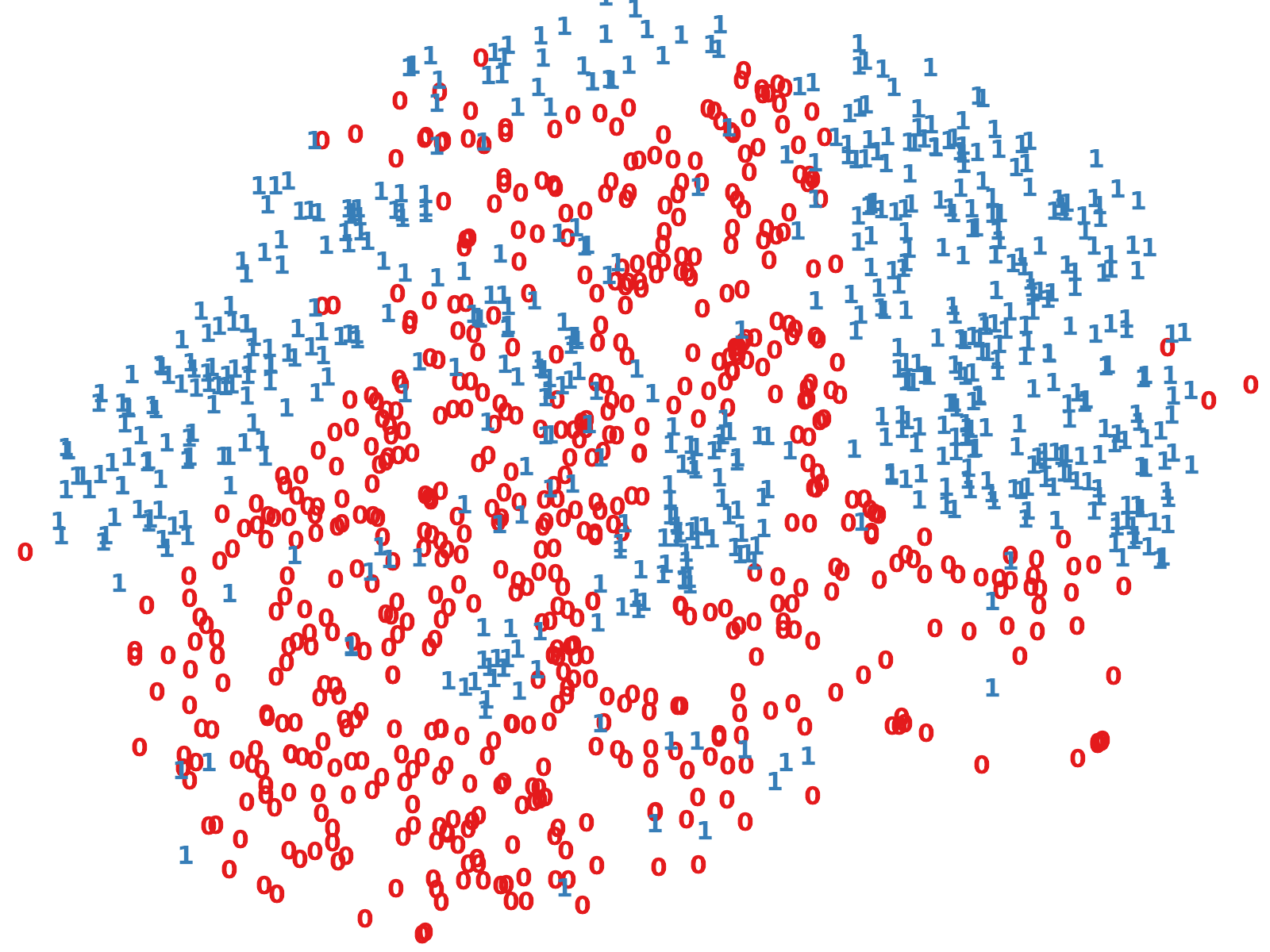}&
\includegraphics[width=0.2\textwidth]{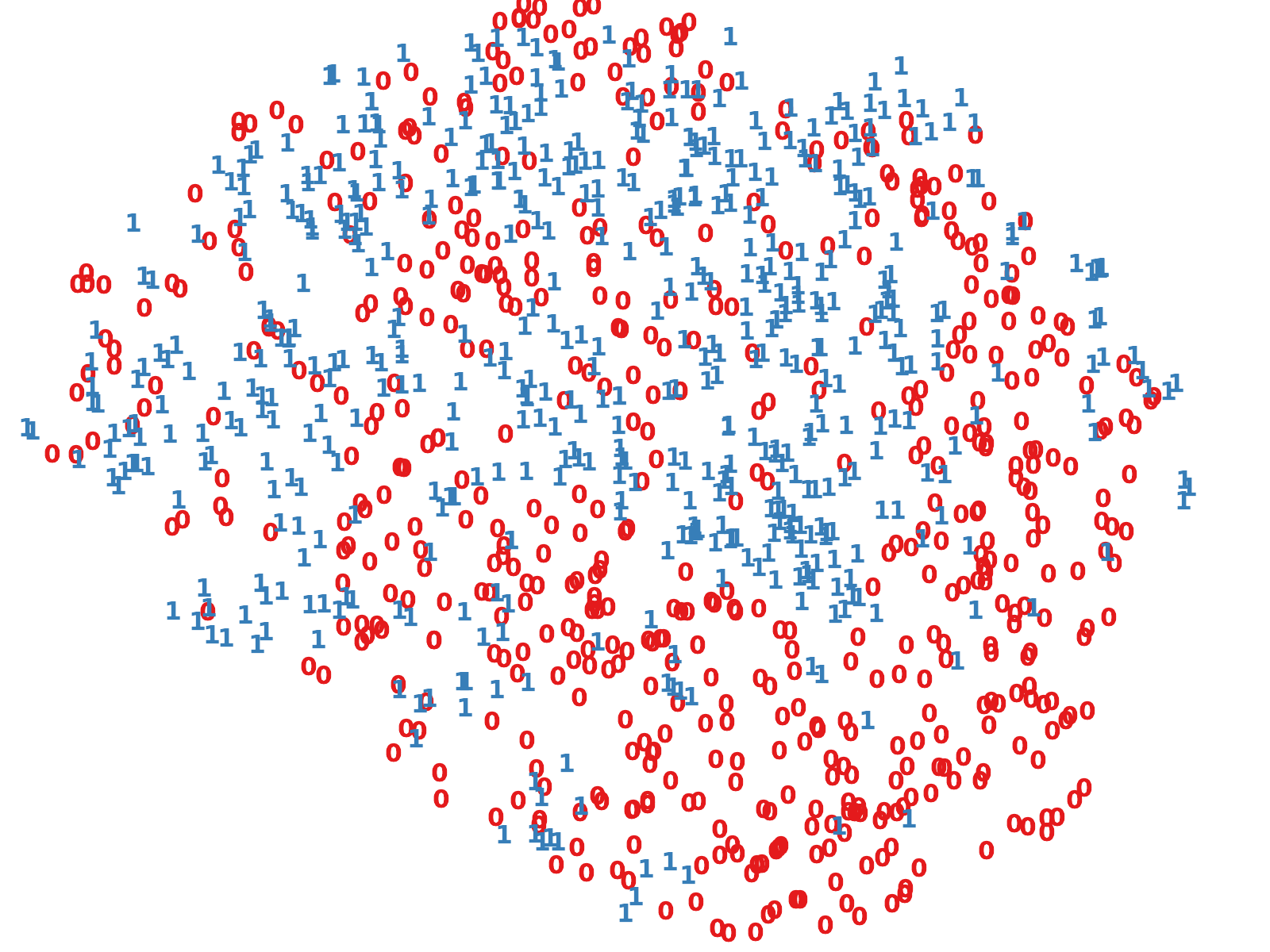}\\
\end{tabular}{}
\caption{t-SNE visualization of Alexnet baseline (left) and our proposed E2E-CE with MCT models (right). Different colors represent different datasets (blue: CheXpert, red: MIMIC). It can be seen that embeddings are well separated in baseline method, suggesting bias are heavily present, while embeddings in our proposed method are more mixed.}
\label{t-SNE-figure}
\end{center}{}
\end{figure}

We plot the feature representation of baseline model and our proposed methods by using t-SNE visualization \cite{maaten2008visualizing} for better demonstration. By looking at Fig. \ref{t-SNE-figure}, we can observe that the dataset bias is clearly present in the vanilla baseline model. On the other hand, the t-SNE embeddings of different datasets are mixed together in our proposed methods, indicating the effectiveness of bias mitigation of our model.
\vspace{-10pt}

\vspace{-15pt}
\subsection{Grad-CAM}
\begin{figure*}[h!]
\begin{center}
%\centerline{\includegraphics[width=\textwidth]{./figures/vqa_compare.jpg}}
\begin{tabular}{ccc}
{\small Original Image} & {\small Grad-CAM of Alexnet} & {\small Grad-CAM of Our Model} \\
\hline
\\
\multicolumn{3}{c}{}\\
\includegraphics[width=0.3\textwidth]{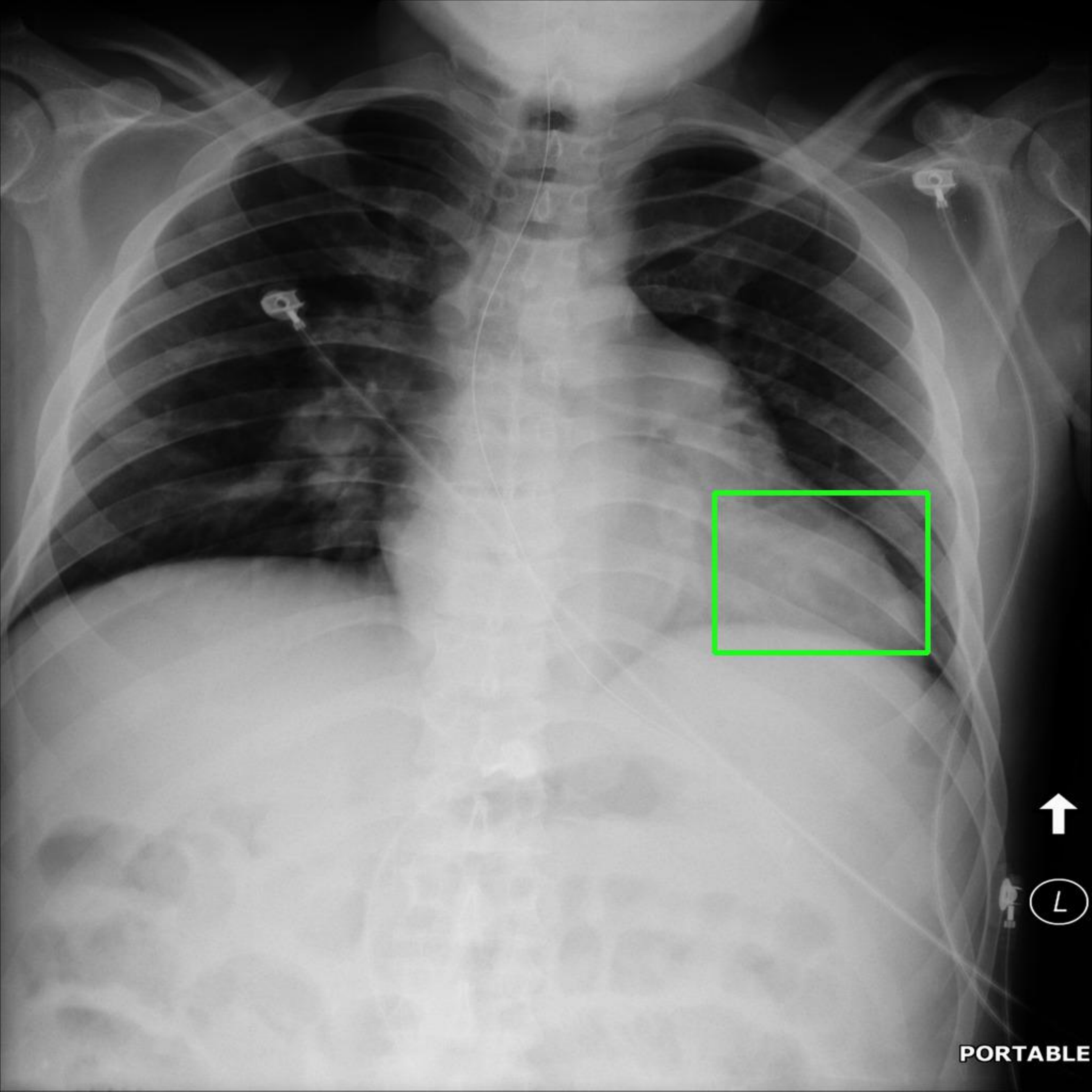}
& \includegraphics[width=0.3\textwidth]{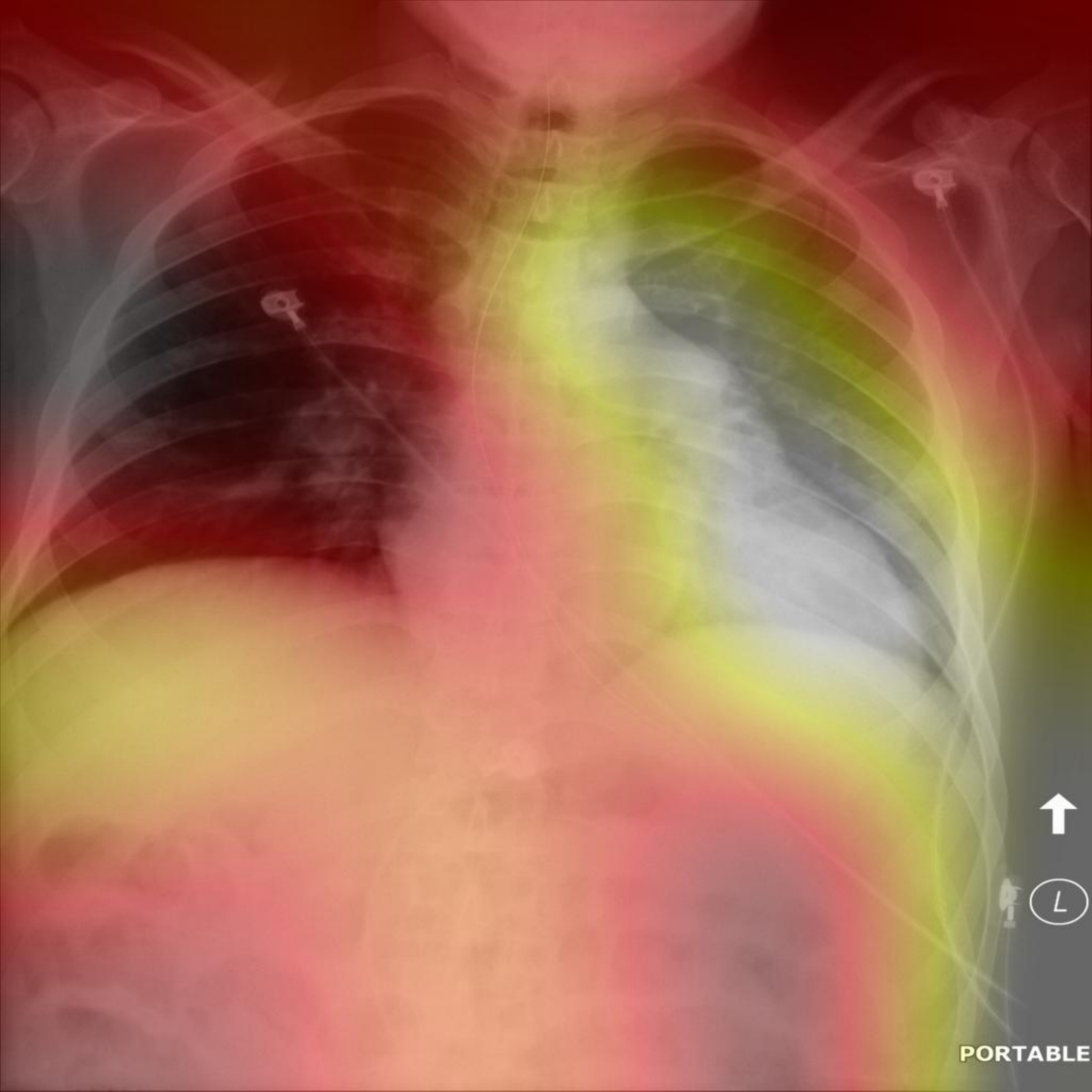}
& \includegraphics[width=0.3\textwidth]{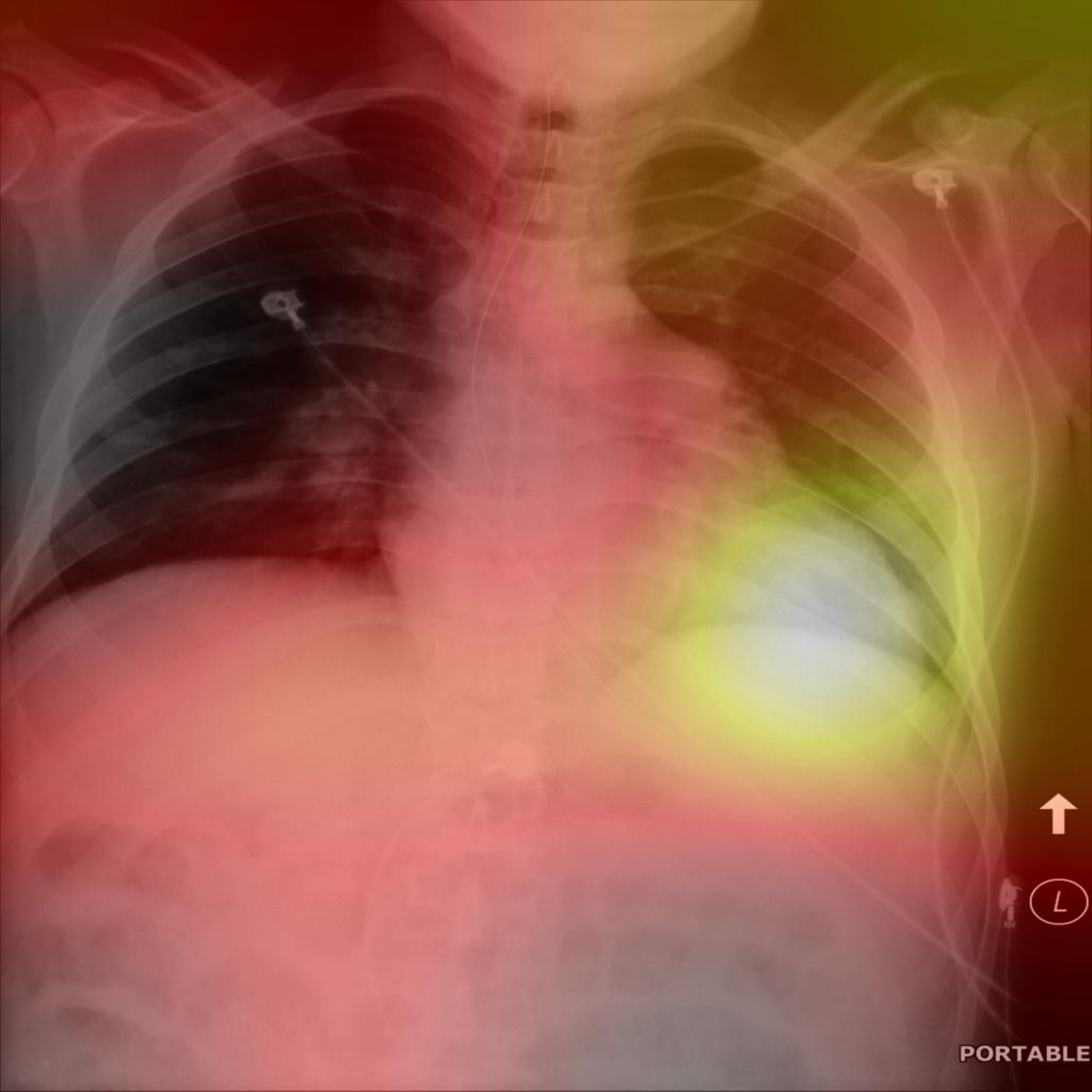} \\
\\
\multicolumn{3}{c}{}\\
\includegraphics[width=0.3\textwidth]{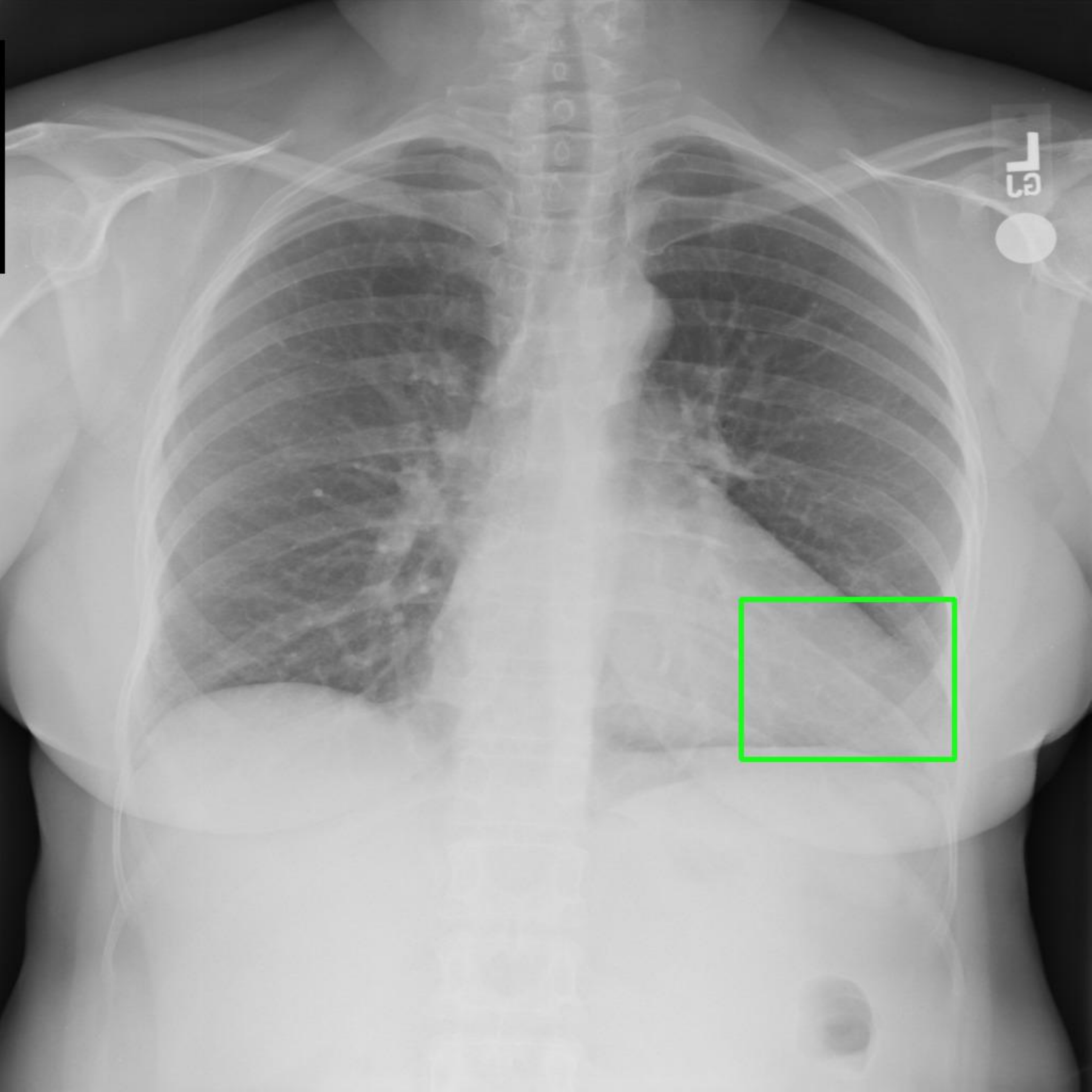}
& \includegraphics[width=0.3\textwidth]{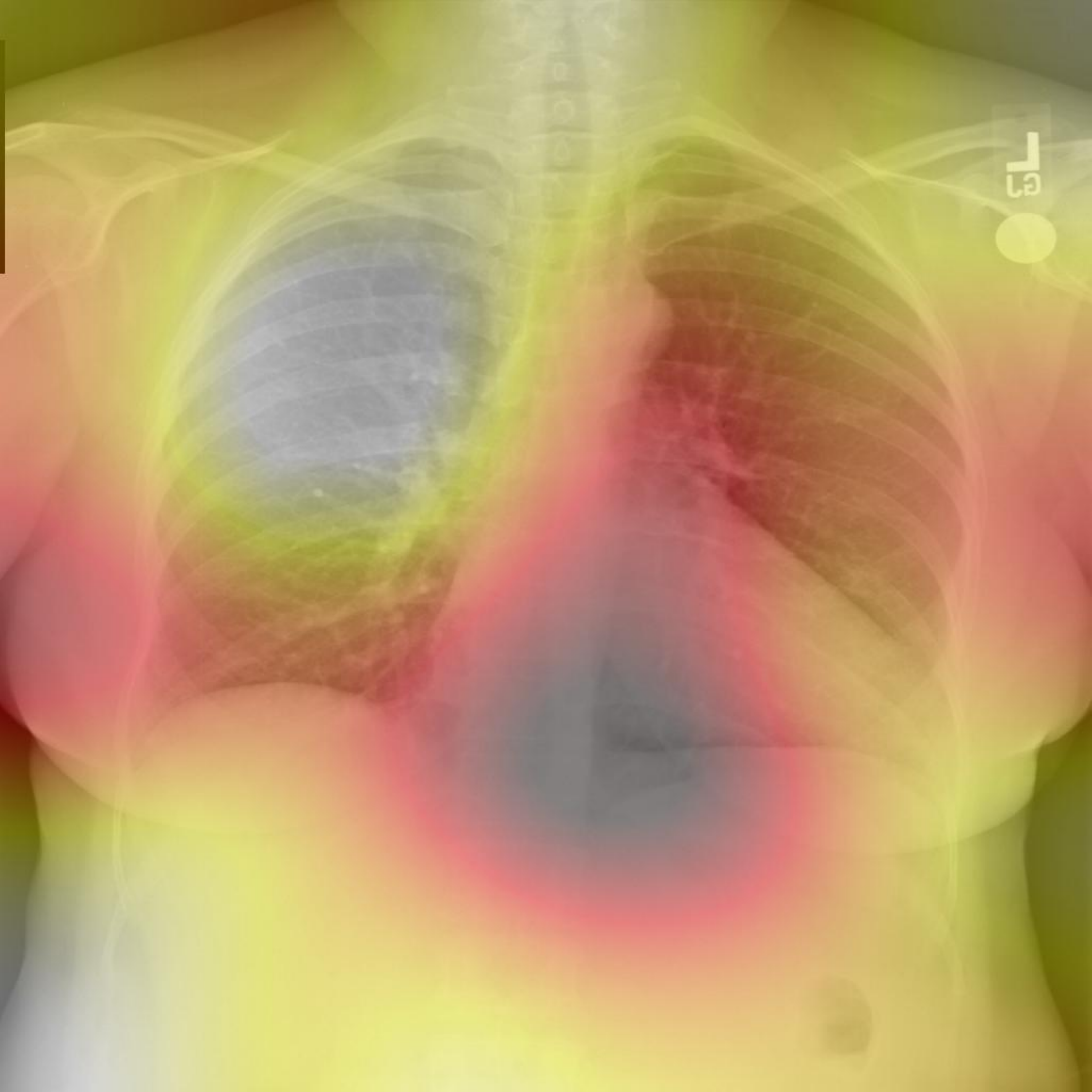}
& \includegraphics[width=0.3\textwidth]{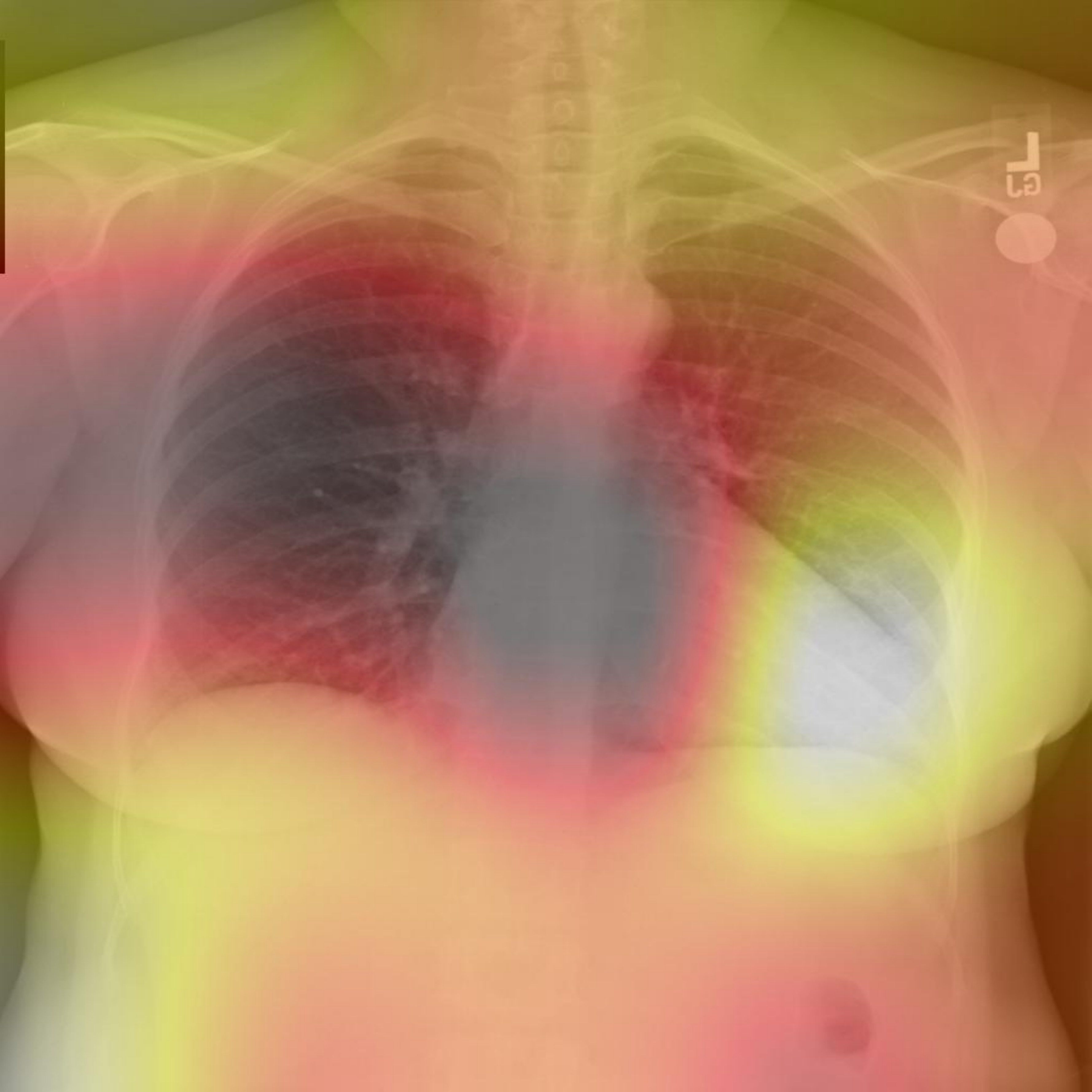} \\
\\
\end{tabular}
\includegraphics[width=1.0\textwidth]{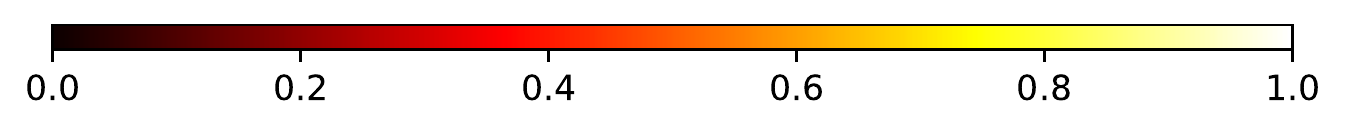}\\
\caption{Grad-CAMs of Alexnet baseline model and our proposed model for Cardiomegaly (enlarged heart) classification. From left to right is original CXR scans, Grad-CAMs of Alexnet overlaid on the original image and Grad-CAMs of our proposed E2E-CE+MCT model, respectively. The green boxes in the original CXRs are the lesion regions labeled by radiologists, whereas in the Grad-CAM figures brighter region indicates higher contribution to the prediction.}
\label{fig:grad-cam}
\end{center}
\end{figure*}

To understand how the bias affect our disease prediction, we plot the gradient-guided activation heatmap (Grad-CAM) \cite{selvaraju2017grad}\cite{uozbulak_pytorch_vis_2019} for the proposed models. Figure \ref{fig:grad-cam} visualizes two sets of results randomly chosen from the external set. We can see that the vanilla Alexnet may take advantage of the unrelated tags and is subject to noise, while our proposed model is being more discriminative and robust.

We also provide quantitative measure for our generated heatmaps by comparing it to the ground-truth annotation using Spearman's rank correlation coefficients \cite{myers2013research}, which measures the rank order between two sets. The results are shown in Table \ref{table:bias_measure}. It can be seen that our proposed method has better correlation with human annotations.

\section{Discussion and Conclusion}\label{discuss}
Due to heterogeneous and confounding data creation process, dataset bias is a common problem in medical datasets. In this study, we showed that (1) naively training and using deep models on medical data is subject to dataset bias, leading to poor generalization ability; (2) our proposed MCT with Bias-regularized Learning framework effectively utilizes multi-source training data, mitigating the damage of dataset bias and closing the performance gap between internal domains and unseen domains without retraining or domain-adaptation. Although we performed case study mainly on medical images, it is easy to adapt it to other forms of biomedical data such as electronic health records, since our proposed methods mainly work on the hidden space of the deep models. Our future work includes investigating the choice of augmentation layer set $\mathbb{S}$ for different neural network architectures and testing our model on other tasks.

% if have a single appendix:
%\appendix[Proof of the Zonklar Equations]
% or
%\appendix  % for no appendix heading
% do not use \section anymore after \appendix, only \section*
% is possibly needed

% use appendices with more than one appendix
% then use \section to start each appendix
% you must declare a \section before using any
% \subsection or using \label (\appendices by itself
% starts a section numbered zero.)
%

\appendices
\section{}\label{conf_mat}
Fig. \ref{fig_name} shows the classification results on a random subset with a simple CNN. The network architecture is simply stacks of 5 convolution blocks followed by a fully-connected layer, with each block to be \textit{conv}-\textit{relu}-\textit{maxpool}. We found that deep CNN could easily differentiate which hospital the data came from, indicating the presence of dataset bias.

\begin{figure}[!t]
\centering
\includegraphics[width=0.3\textwidth]{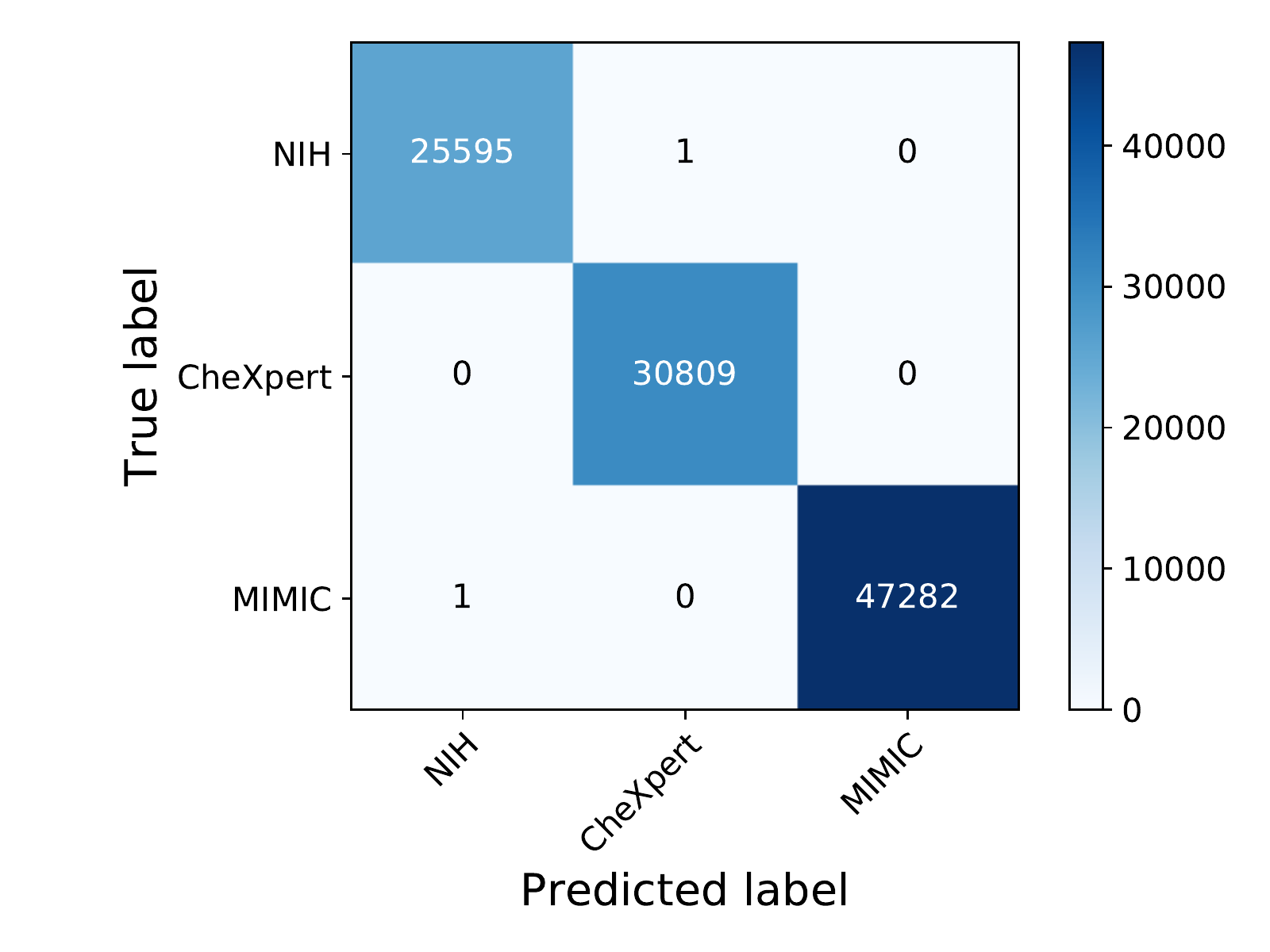} 
%where an .eps filename suffix will be assumed under latex, 
% and a .pdf suffix will be assumed for pdflatex; or what has been declared
% via \DeclareGraphicsExtensions.    

\caption{Confusion matrix of dataset classification results. A near-perfect accuracy is obtained, suggesting that dataset bias are present in the three datasets (MIMIC, CheXpert and NIH).}
\label{fig_name}
\end{figure}

\section{}\label{toy exp}
We explicitly test our novel MCT domain-guided augmentation methods on three popular domain-generalization datasets. They are
\begin{itemize}
    \item Google Fonts \cite{shankar2018generalizing}: the task is to classify 36 characters collected from 109 fonts;
    \item Rotated-MNIST \cite{Ghifary_2015_ICCV}: this dataset is created by rotating the original MNIST dataset with 6 different degress: 0, 15, 30, 45, 60 and 75. Each image now has a digit label and rotation angle as its domain;
    \item Office-Caltech \cite{gong2012geodesic}: there are in total ten different common object categories of four domains (Amazon, Caltech, DSLR and Webcam). Domains tend to have very different viewing angles, object background and etc. 
\end{itemize}{}

The experiment details are as follows: for Google Fonts and Rotated-MNIST (R-MNIST), We use the same configuration as in \cite{shankar2018generalizing}; for Office-Caltech, we use the same setting as Google Fonts. The selected layers $\mathbb{S}$ for the three experiments are all the layers including the input but without the first convolution and the last dense layer. Results are shown on Table \ref{table:toy}, where the baseline method of Google Fonts and Office-Caltech is LeNet \cite{lecun1998gradient} without special training, while that of R-MNIST is CCSA \cite{motiian2017unified}. Our proposed MCT augmentation method surpasses other comparing model by a large margin. 

\begin{table}[!ht]
% increase table row spacing, adjust to taste
\renewcommand{\arraystretch}{1.3}
% if using array.sty, it might be a good idea to tweak the value of
% \extrarowheight as needed to properly center the text within the cells
\caption{Experiments on Toy Dataset}
\label{table:toy}
\centering
%% Some packages, such as MDW tools, offer better commands for making tables
%% than the plain LaTeX2e tabular which is used here.
\begin{threeparttable}
\begin{tabular}{|c|c|c|c|}
\hline
 Models & Fonts & R-MNIST & Office-Caltech \\
\hline
Baseline & 68.5\footnotemark[1]  & 95.6\footnotemark[1]  & 41.6 \\
\hline
DANN \cite{ganin2016domain}  & 68.9\footnotemark[1]  & 98.0\footnotemark[1]  & 43.8 \\
\hline
CrossGrad \cite{shankar2018generalizing} & 72.6\footnotemark[1]  & 98.6\footnotemark[1]  & 44.3  \\
\hline
MCT (ours) & 73.1 & 99.6 & 47.8  \\
\hline
\end{tabular}

\begin{tablenotes}
        \footnotesize
        \item[1] Results are taken directly from \cite{shankar2018generalizing}
\end{tablenotes}
\end{threeparttable}
\end{table}
%\clearpage
%\footnotetext[1]{Results are taken directly from \cite{shankar2018generalizing}}
% you can choose not to have a title for an appendix
% if you want by leaving the argument blank

\iffalse
% use section* for acknowledgment
\section*{Acknowledgment}

The authors would like to thank...
\fi

% Can use something like this to put references on a page
% by themselves when using endfloat and the captionsoff option.
\ifCLASSOPTIONcaptionsoff
  \newpage
\fi

% trigger a \newpage just before the given reference
% number - used to balance the columns on the last page
% adjust value as needed - may need to be readjusted if
% the document is modified later
%\IEEEtriggeratref{8}
% The "triggered" command can be changed if desired:
%\IEEEtriggercmd{\enlargethispage{-5in}}

% references section

% can use a bibliography generated by BibTeX as a .bbl file
% BibTeX documentation can be easily obtained at:
% http://mirror.ctan.org/biblio/bibtex/contrib/doc/
% The IEEEtran BibTeX style support page is at:
% http://www.michaelshell.org/tex/ieeetran/bibtex/
%\bibliographystyle{IEEEtran}
% argument is your BibTeX string definitions and bibliography database(s)
%\bibliography{IEEEabrv,../bib/paper}
%
% <OR> manually copy in the resultant .bbl file
% set second argument of \begin to the number of references
% (used to reserve space for the reference number labels box)
\bibliographystyle{IEEEtran}
\bibliography{egbib}
\iffalse
%\begin{thebibliography}{1}
%\bibitem{IEEEhowto:kopka}
%H.~Kopka and P.~W. Daly, \emph{A Guide to \LaTeX}, 3rd~ed.\hskip 1em %plus
%  0.5em minus 0.4em\relax Harlow, England: Addison-Wesley, 1999.
%\end{thebibliography}

% biography section
% 
% If you have an EPS/PDF photo (graphicx package needed) extra braces are
% needed around the contents of the optional argument to biography to prevent
% the LaTeX parser from getting confused when it sees the complicated
% \includegraphics command within an optional argument. (You could create
% your own custom macro containing the \includegraphics command to make things
% simpler here.)
%\begin{IEEEbiography}[{\includegraphics[width=1in,height=1.25in,clip,keepaspectratio]{mshell}}]{Michael Shell}
% or if you just want to reserve a space for a photo:

\begin{IEEEbiography}{Michael Shell}
Biography text here.
\end{IEEEbiography}

% if you will not have a photo at all:
\begin{IEEEbiographynophoto}{John Doe}
Biography text here.
\end{IEEEbiographynophoto}

% insert where needed to balance the two columns on the last page with
% biographies
%\newpage

\begin{IEEEbiographynophoto}{Jane Doe}
Biography text here.
\end{IEEEbiographynophoto}
\fi

% You can push biographies down or up by placing
% a \vfill before or after them. The appropriate
% use of \vfill depends on what kind of text is
% on the last page and whether or not the columns
% are being equalized.

%\vfill

% Can be used to pull up biographies so that the bottom of the last one
% is flush with the other column.
%\enlargethispage{-5in}

% that's all folks
\end{document}